\documentclass[sigplan,authorversion]{acmart}

\setcopyright{acmcopyright}

\usepackage{booktabs}   
\usepackage{subcaption} 
\usepackage{listings}
\usepackage{balance}
\usepackage{multirow}
\usepackage[utf8]{inputenc}

\usepackage[linesnumbered,ruled,vlined]{algorithm2e}

\usepackage{cleveref}

\begin{document}

\title[Supporting OSR in Unstructured Languages]{Supporting On-Stack Replacement in Unstructured Languages by Loop Reconstruction and Extraction}
\titlenote{This research project is partially funded by Oracle Labs.}        


\author{Raphael Mosaner}
\affiliation{
	\institution{Johannes Kepler University Linz}
	\country{Austria}
}\email{raphael.mosaner@jku.at} 

\author{David Leopoldseder}
\affiliation{
	\institution{Johannes Kepler University Linz}
	\country{Austria}
}\email{david.leopoldseder@jku.at} 

\author{Manuel Rigger}
\affiliation{
	\institution{ETH Zürich}
	\country{Switzerland}
}\email{manuel.rigger@inf.ethz.ch} 

\author{Roland Schatz}
\affiliation{
	\institution{Oracle Labs}
	\country{Austria}
}\email{roland.schatz@oracle.com}

\author{Hanspeter Mössenböck}
\affiliation{
	\institution{Johannes Kepler University Linz}
	\country{Austria}
}\email{hanspeter.moessenboeck@jku.at}

\begin{abstract}
On-stack replacement (OSR) is a common technique employed by dynamic compilers to reduce program warm-up time.
OSR allows switching from interpreted to compiled code during the execution of this code.
The main targets are long running loops, which need to be represented explicitly, with dedicated information about condition and body, to be optimized at run time.
Bytecode interpreters, however, represent control flow implicitly via unstructured jumps and thus do not exhibit the required high-level loop representation.
To enable OSR also for jump-based---often called unstructured---languages, we propose the 
partial reconstruction of loops in order to explicitly represent them in a bytecode interpreter.
Besides an outline of the general idea, we implemented our approach in Sulong, a bytecode 
interpreter for LLVM bitcode, which allows the execution of C/C++. 
We conducted an evaluation with a set of C benchmarks, which showed speed-ups in warm-up of up to 9x for certain benchmarks.
This facilitates execution of programs with long-running loops in rarely called functions, which would yield significant slowdown without OSR.
While shown with a prototype implementation, the overall idea of our approach is generalizable 
for all bytecode interpreters.
\end{abstract}

\copyrightyear{2019}
\acmYear{2019}
\acmConference[MPLR '19]{Proceedings of the 16th ACM SIGPLAN International Conference on Managed Programming Languages and Runtimes}{October 21--22, 2019}{Athens, Greece}
\acmBooktitle{Proceedings of the 16th ACM SIGPLAN International Conference on Managed Programming Languages and Runtimes (MPLR '19), October 21--22, 2019, Athens, Greece}
\acmPrice{15.00}
\acmDOI{10.1145/3357390.3361030}
\acmISBN{978-1-4503-6977-0/19/10}



\maketitle
\renewcommand{\shortauthors}{R. Mosaner, D. Leopoldseder, M. Rigger, R. Schatz, H. Mössenböck}

\section{Introduction}
On-stack replacement~\cite{Fink2003} is a technique employed by dynamic compilers for reducing program warm-up time~\cite{Barrett2017}. 
Based on Barrett's work~\cite{Barrett2017}, we define program warm-up as the time it takes a dynamic compiler to identify and compile the hot parts of a program to reach a steady state of peak performance.
OSR usually works by detecting hot but not yet compiled code and performing a switch from the interpreted to the compiled version of this code while it is being executed.
This is most useful for long-running loops, which can take up most of the execution time and should be compiled as soon as possible.
However, method-based compilation systems do not directly support OSR, because methods are the most fine-grained compilation units.
For instance, a computation intensive loop in a \textit{main}-function would never be compiled, as the \textit{main}-function is only called once and thus never considered hot.
To make OSR work in such systems, it would be possible to extract loops into separate functions which can be independently compiled by a method-based compiler even when it lacks support for OSR.
While for structured languages, loop bodies could be extracted to separate functions to allow for their quick compilation, unstructured languages lack high-level loop information. 
To tackle this issue and support OSR for unstructured languages in method-based compilation systems we propose an approach in which we (1) reconstruct loop information from unstructured, basic block-based program representations and (2) explicitly represent loops as separate functions to enable OSR.
\par 
We implemented our approach in Sulong~\cite{Rigger2016,Rigger2018}, an interpreter for LLVM IR, which suffers from long warm-up time~\cite{Rigger2018}.
LLVM IR is an unstructured language that does not explicitly represent loops.
Sulong is used in the multi-lingual GraalVM~\cite{Duboscq2013,Stadler2014,Leopoldseder2018} to implement native function interfaces of dynamic languages such as TruffleRuby~\cite{Daloze2016}.
All these language implementations are based on a common language implementation framework, called Truffle~\cite{Wurthinger17}.
It uses dynamic-compilation based on profiling information gathered during interpretation of 
abstract syntax trees (ASTs) for efficient execution of the implemented language.
Truffle does not directly support OSR.
However, it provides language implementers with a framework to enable loop-level OSR which requires extracting loops to form separate compilation units as outlined above.
We demonstrate that reconstructing loops and extracting them to separate units gives significant speed-ups.
Specifically, our evaluation with a set of well-known benchmarks shows significant reductions by up to 9x in program warm-up given that OSR is applicable.
Note that our approach can be used by other Truffle bytecode-based implementations lacking high-level loop information, including GraalSqueak~\cite{Niephaus18}, Truffle Java~\cite{Grimmer2017} and Truffle CIL.
Furthermore, it is applicable in any compiler with a background system which provides means for establishing mappings between extracted control flow.

In summary, this paper contributes the following:

\begin{itemize}
	\item a novel multi-tier approach for employing OSR for unstructured languages
	\begin{itemize}
		\item detection of loops from unstructured control flow
		\item reconstruction of high-level loops
		\item extraction of loops into separate functions to model loop-level OSR using method-based compilation
	\end{itemize}
	\item a prototype implementation in Sulong,
	\item extensive experiments suggesting significant improvements in warm-up performance.
\end{itemize}

\section{Background}
This section provides the necessary background information to understand our approach for supporting OSR in Truffle-based implementations of unstructured languages.
We first give an overview of the Truffle language implementation framework and its OSR mechanism for structured languages.
We then discuss bytecode-interpreter-based Truffle languages and the problems they cause for OSR.
We implemented a prototype of our approach in Sulong, an interpreter for LLVM IR.
Thus, we also give background information about Sulong and LLVM.

\subsection{Truffle and Graal}
	The Truffle language implementation framework \cite{Wurthinger2013} provides language implementers with means for creating efficient Abstract Syntax Tree (AST) interpreters.
    In an AST interpreter, each operation is implemented by a node that computes its value by potentially evaluating its children, to then return this value to its parent.
    Truffle uses the dynamic Graal compiler~\cite{Wurthinger2013} to efficiently compile frequently-executed functions to machine code.
    When invoked, it recursively inlines all node execution methods of the AST (which is a form of partial evaluation ~\cite{Futamura1999}), to then further optimize it.
  
    To achieve optimal performance, language implementers have to use various constructs of the Truffle framework.
    Graal, when used as a standard Java compiler, is capable of performing OSR on loop-level.
    Truffle, however, provides an interface that needs to be implemented by guest language nodes to support OSR for structured languages.
    This \texttt{RepeatingNode} interface assumes that the interpreted language has a concept of high-level loops in order to enable OSR.  

    \sloppy{}
    \Cref{lst:executeRepeating} shows a typical implementation of a \texttt{RepeatingNode}.
    The \texttt{executeRepeating()} method executes only a single loop iteration and returns either \texttt{true} if the loop is to be executed again or \texttt{false} if it was the loop's last iteration.
    The Truffle framework wraps the \texttt{RepeatingNode} with a \texttt{LoopNode} that executes the loop.
    Its implementation is transparent to the language implementer.
    Since each loop iteration is performed individually, the \texttt{LoopNode} can trigger compilation after any loop iteration and switch from interpreted to compiled execution. 
    Currently, Truffle uses a constant iteration threshold before OSR-compiling a loop.
    Note that this technique is generally applicable to compilation systems for adding OSR support.	
\begin{lstlisting}[caption={A typical implementation of a \texttt{RepeatingNode} for structured languages. Both \texttt{conditionNode}(loop condition) and \texttt{bodyNode}(loopBody) are child nodes of the \texttt{RepeatingNode}. Execution of the body can be interrupted by \texttt{break}- or \texttt{continue}-statements, which throw an exception.},label=lst:executeRepeating, frame=bt, language=Java,float]
class LLVMRepeatingNode implements RepeatingNode{
  public boolean executeRepeating(Frame frame) {
    if ((boolean)conditionNode.execute(frame)) {
      try{
        bodyNode.execute(frame);
      } catch (BreakException e) {
        return false;
      } catch (ContinueException e) {
        return true;
      }
      return true;
    }
    return false;
  }
}
\end{lstlisting}
\subsection{Sulong and LLVM IR}
Sulong \cite{Rigger2016,Rigger2018} is a bytecode interpreter for LLVM IR built on top of Truffle.
LLVM IR is the intermediate representation of the LLVM compilation framework~\cite{Lattner2004}, which provides front ends for various languages such as C/C++ and Fortran, allowing Sulong to execute these languages via LLVM IR.
Unlike structured languages, LLVM IR has no notion of loops; instead, it uses basic blocks~\cite{Torczon2007}, which consist of sequential instructions that end with a terminating instruction transferring control flow to another basic block.
Since these terminating instructions are conditional or unconditional jumps, LLVM IR is considered a unstructured language~\cite{Dijkstra1972}.
Furthermore, it can exhibit irreducible control flow ~\cite{Erosa1994}, where loops have more than one header (for example by having a jump from outside into the loop).
This prevents loops from being representable by a single AST node without further processing them, to handle the irreducibility.

\Cref{lst:problemSubFigC} shows the LLVM IR of the \texttt{main} function from \Cref{lst:problemSubFigA} with basic block 1 forming the loop.
The control-flow graph of basic blocks can be inferred from \Cref{fig:problemSubFigD}.
Basic block 0 is the function start which unconditionally jumps to basic block 1, which is the loop header.
This block is then terminated with a conditional branch instruction that either transfers control back to itself or exits the loop by jumping to basic block 2, which has only the \texttt{ret} instruction for returning from the function.
In this simple example, the loop consists of only one block, which holds both, condition and body. \par
Sulong's execution mode is different from classical AST-based Truffle language implementations, such as TruffleRuby or GraalPython.
It uses a \texttt{BlockDispatchNode} that dispatches the control flow between the individual LLVM IR \texttt{BasicBlockNodes}~\cite{Rigger2016}, whereas the high-level control flow is unknown to Sulong (see \Cref{fig:problemSubFigD}).
This makes Sulong's structure similar to a bytecode interpreter (see \Cref{lst:interpreterloop}).
For each basic block, Sulong builds a conventional AST with a \texttt{BasicBlockNode} as its root.
The \texttt{BasicBlockNode} executes all its instructions sequentially.
The last instruction of the basic block returns an index that is used to fetch the next basic block to be executed.
To return from a function, \texttt{-1} is returned as a special value.
GraalSqueak~\cite{Niephaus18} for example, is implemented in a similar way.

\subsection{Partial Evaluation of Bytecode Interpreters} \label{sec:PE}
Truffle optimizes functions at run time by partially evaluating their ASTs.
Partial evaluation (PE) recursively inlines the execution code of children AST nodes into their parent node until one final compilation unit is derived.
However, Sulong represents loops with its bytecode-interpreter-like basic block dispatch node (see \Cref{lst:interpreterloop}).
In the case of loops, the standard version of PE would infinitely inline successor blocks of already inlined blocks.
Thus, the partial evaluation had to be adapted~\cite{Rigger2016} to support bytecode interpreters as described below.

Sulong's basic block dispatch loop is treated in a special way by Graal.
As for AST-based Truffle interpreters, Graal starts to partially evaluate the interpreter, starting with the first basic block.
The basic blocks and the indices of their possible successor blocks are constant during run time.
Thus, Graal can continue to recursively inline all successor blocks.
In contrast to AST-based approaches, however, Graal keeps track of paths that have already been expanded before, which it can determine based on their index in the constant basic block array.
If it detects an already partially evaluated block, Graal connects the path from the currently processed block with the previously expanded successor block.
Effectively, it reconstructs guest-language-level loops.
The merging of already expanded basic blocks also effectively limits loop expansion to the 
number of original blocks in the LLVM IR program.
Even though bytecode might exhibit irreducible control flow, Graal does not support it.
Reducible control flow yields easier optimization properties and faster algorithms.
Thus, when encountering irreducible control flow, Graal, wraps the irreducible blocks in a dispatch loop during partial evaluation, as it is also done in Sulong.

\begin{lstlisting}[caption={Block Dispatch Loop; compacted for brevity},label=lst:interpreterloop, frame=bt, language=Java, float]
int blockIndex = 0
while(blockIndex!=-1){
  blockIndex = basicBlocks[blockIndex].execute();
}
\end{lstlisting}

\subsection{Problem of OSR for Bytecode Interpreter Languages}
\begin{figure*}
	\centering
	\begin{subfigure}[l]{.45\textwidth}
		\centering
\begin{lstlisting}[frame=bt, language=Java]
int main(int argc, char** argv) {
int i = 0;
do {
processRequest();
i++;
} while (i < 1000000);

return 0;
}
\end{lstlisting}
		\caption{Source level function}
		\label{lst:problemSubFigA}
	\end{subfigure}%
	\begin{subfigure}[c]{.45\textwidth}
		\centering
		\includegraphics[width=0.8\linewidth]{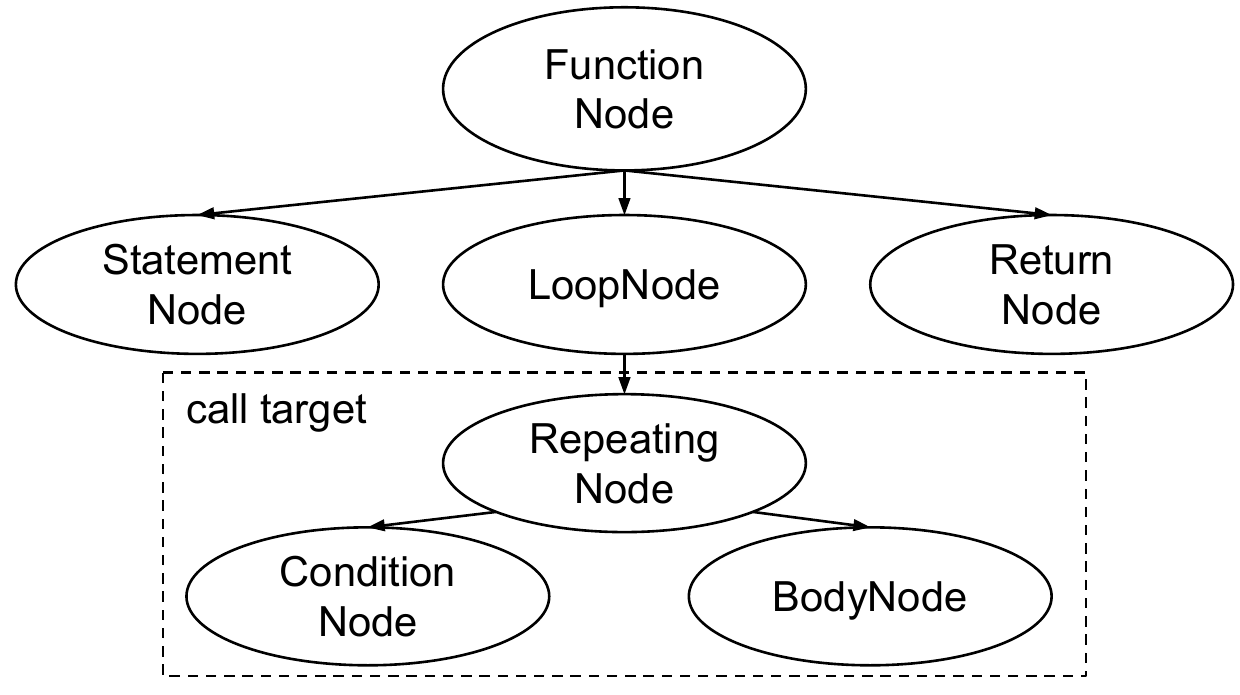}
		\caption{Truffle AST of program with high-level loop}
		\label{fig:problemSubFigB}
	\end{subfigure}
	\begin{subfigure}[c]{.45\textwidth}
		\centering
\begin{lstlisting}[frame=bt, basicstyle=\scriptsize, language=llvm]
define i32 @main(i32 %argc, i8** %argv) #0 {
br label %1  (basic block 0)

; <label>:1  (basic block 1)
%i.0 = phi i32 [ 0, %0 ], [ %2, %1 ]
call void @processRequest()
%2 = add nsw i32 %i.0, 1
%3 = icmp slt i32 %2, 1000000
br i1 %3, label %1, label %4

; <label>:4  (basic block 2)
ret i32 0
}
\end{lstlisting}
		\caption{Function in LLVM IR}
		\label{lst:problemSubFigC}
	\end{subfigure}%
	\begin{subfigure}[c]{.45\textwidth}
		\centering
		\includegraphics[width=0.8\linewidth]{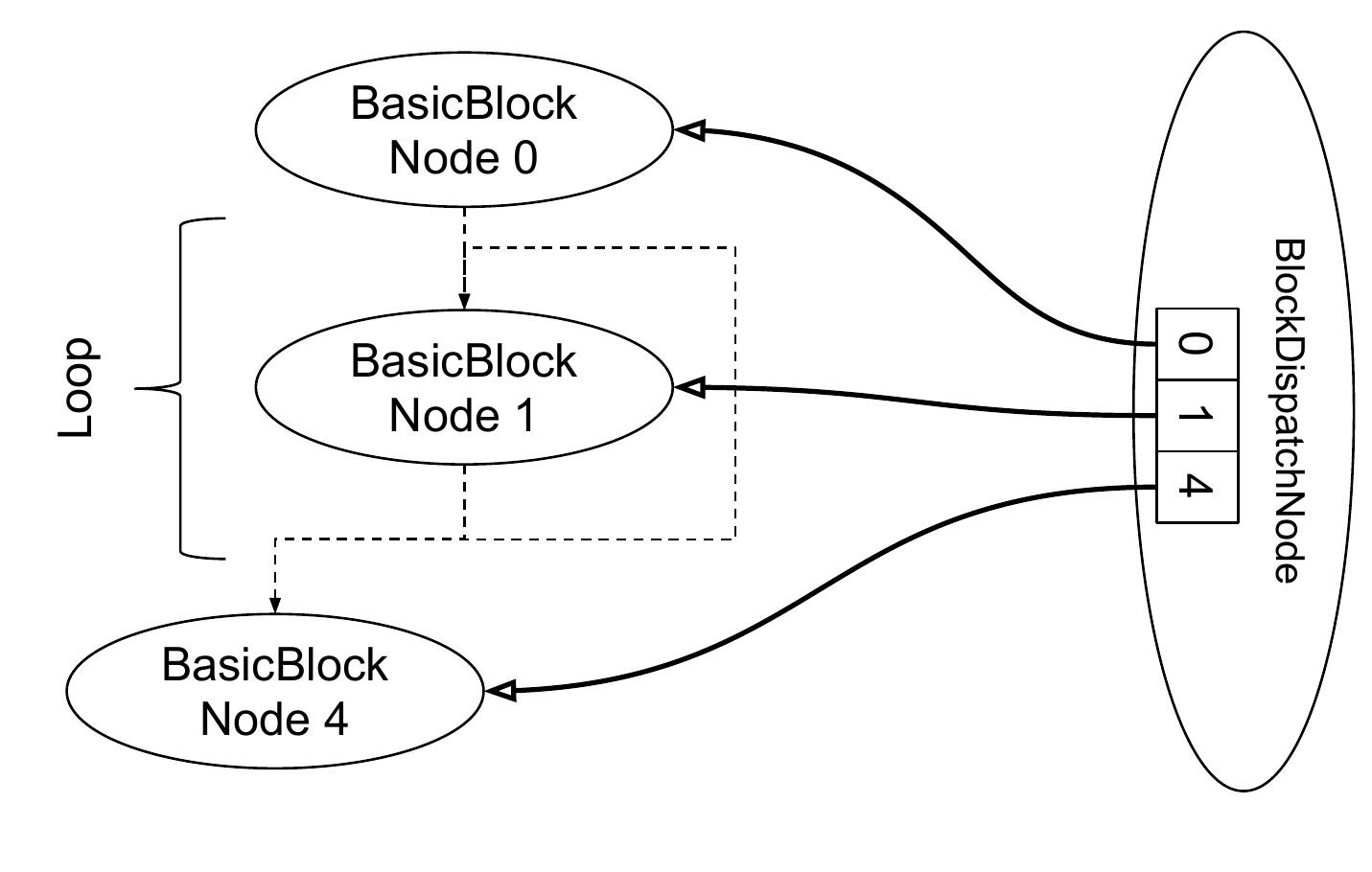}
		\caption{Bytecode interpreter block dispatch; control flow is depicted as a dashed line}
		\label{fig:problemSubFigD}
	\end{subfigure}%
	\caption{Test program: Source code; LLVM IR, Truffle AST and basic block interpreter.}
	\label{fig:problemStatement}
	\vspace{10pt}
\end{figure*}
	The difference between conventional AST- and bytecode-interpreted languages also manifests itself in how OSR can be applied. 
	In order to switch from an interpreted to a compiled loop, a mapping between the frame state of the interpreter and the compiled code must be established	to continue execution in the same state after the transition~\cite{Fink2003}.
	The frame state describes all dynamic information of the current program execution including local variables and the program counter.
	\Cref{lst:problemSubFigA} depicts a source program with structured control flow, which is translated into a Truffle AST as shown in \Cref{fig:problemSubFigB}. 
	The \texttt{RepeatingNode} wraps one loop iteration and is turned into a \texttt{CallTarget}, which is equivalent to creating a special function for executing the loop body. 
	Thus, when OSR is triggered, the mapping of the program state between interpreted and compiled loop is trivial, as only the frame state at the loop's \texttt{CallTarget} invocation has to be mapped to the state at the beginning of the compiled loop.
	Hence, the frame state between two loop iterations is used as a parameter for the loop \texttt{CallTarget}, which suffices to continue execution in the compiled loop. 
	
	However, in bytecode interpreter languages, like LLVM IR, there is no concept of loops because not all bytecodes are directly derived from structured languages. 
	\Cref{lst:problemSubFigC} shows the structure of basic blocks derived after compiling the \texttt{main} function from \Cref{lst:problemSubFigA} to LLVM IR.
	The \texttt{BlockDispatchNode}, which is the AST root node for functions in Sulong, stores the array of basic blocks without any knowledge of the high-level control flow. This is visualized in \Cref{fig:problemSubFigD}.
	Therefore, even with the compiled versions of the two programs (structured C program and LLVM IR code) being identical after partial evaluation, the interpreted loop cannot be mapped to the compiled loop, preventing Truffle's OSR mechanism to be applied for bytecode interpreter languages.

\section{OSR for Bytecode-based Truffle Interpreters}
	
	In this section, we present our approach to enable OSR for bytecode-based Truffle interpreters. 
	The idea is general enough to be applicable to any bytecode-based language.
	For this paper we use our prototype implementation in Sulong as an example.
	To re-use Truffle's built-in mechanism for OSR, we reconstruct loops from the potentially unstructured control flow and create new nodes for representing high-level loops according to the Truffle interface.\par
	Below we list all necessary steps to support OSR in Sulong:
    \begin{enumerate}
    	\item The language implementation's parser needs to detect loops in the function before executing it the first time.
    	\item After the loops are identified, the parser also needs to determine loop relations to handle nested loops separately.
    	\item Then, the Truffle implementation needs to create \texttt{LoopNodes}, and integrate them in the block dispatch loop such that execution of the next \texttt{BasicBlockNode} after a \texttt{LoopNode} works seamlessly.
	    \item Finally, during execution, the \texttt{LoopNode} needs to communicate the successor of the loop to the enclosing loop or function.
    \end{enumerate}
	We expect language implementations for other unstructured languages to use the same steps to implement support for OSR.	
	The remainder of this section describes each step in detail.

\subsection{Loop Detection} \label{ssec:loopDetection}
\begin{algorithm}
	\caption{{\sc findLoops} detects loops in a CFG of basic blocks}
	\label{algo:loopDetection}
	\DontPrintSemicolon
	
	\KwIn{The basic block $b$ where to start a depth-first loop detection}
	\KwOut{A set of loops found in the graph}
	\If{b.visited} {
		\If{b.active}{
			$makeLoop(b)$	\tcp*{new loop with header b}
			\tcp*{\textit{b.isLoopHeader = true}}
			\Return{b.loops} \tcp*{loops which contain b}
		}
		\ElseIf{b.isLoopHeader}{
			\Return{setSubtraction(b.loops, b.mainLoop)} \tcp*{return all loops except the loop}
			\tcp*{which is associated to the header}
		}
		\Else{
			\Return{b.loops} \tcp*{loops which contain b}
		}
	}
	\;
	$b.visited \gets true$,	$b.active \gets true$
	\;\;
	\ForEach{block $s \in b.successors$} {
		$b.loops.add(findLoops(s))$	\tcp*{add loops}
		\tcp*{propagated from successors}
	}
	\ForEach{loop $l \in b.loops$}{
		$l.add(b)$	\tcp*{add b to containing loops}
	}
	\;
	$b.active \gets false$
	\;\;
	\If{b.isLoopHeader}{
		\Return{setSubtraction(b.loops, b.mainLoop)} \tcp*{return all loops except the loop}
		\tcp*{which is associated to the header}
	}
	\Else{
		\Return{b.loops} \tcp*{loops which contain b}
	}
\end{algorithm}

	When parsing the basic blocks of the LLVM IR, we initially build a control-flow graph (CFG) by identifying predecessors and successors of all blocks, which is needed for the loop detection algorithm.    
	Our loop-detection algorithm is based on Graal's depth-first algorithm~\cite{Wimmer2004} for detecting loops in bytecode is shown in \Cref{algo:loopDetection}.
	
	We now discuss \Cref{algo:loopDetection} in detail.
	It takes basic block \textit{b} as parameter for the next depth-first step.
	A block can be marked as \textit{visited}, which means that it was already seen by the algorithm.
	When a block is marked \textit{active}, the algorithm has started but not finished processing it and the block is still further up in the current recursive traversal.
	 
	The algorithm checks whether \textit{b} is already visited (line 1) and if it is still 
	marked as active (line 2). 
	If both conditions hold, \textit{b} must have been reached via a backedge and thus a loop with header \textit{b} is detected.
	A new loop is created (done by function \texttt{makeLoop} in line 3) with \textit{b} as header.
	This loop is denoted \textit{mainLoop} of \textit{b}.
	The algorithm always returns the set of loops which contain \textit{b}'s predecessor in the traversal which we denote as back-propagation (lines 4, 6, 8, 20 and 22).
	There are three different cases: 
\begin{itemize}
	\item \textit{b} is visited and active and thus is detected to be a loop header (line 4): Then, all of \textit{b}'s loops contain \textit{b}'s predecessor in the recursive traversal (source of the backedge).
	This includes the newly created loop.
	
	\item \textit{b} is visited but not active anymore and is marked as loop header (line 5):
	This means that \textit{b} is not reached via a backedge. 
	Therefore all of \textit{b}'s loops except its \textit{mainLoop} contain \textit{b}'s predecessor in this traversal.
	In the algorithm we remove the \textit{mainLoop} from \textit{b}'s loops using a set subtraction (lines 6 and 20).
	
	\item \textit{b} is visited but not active anymore and no loop header (line 7):
	Therefore, all of \textit{b}'s loops contain \textit{b}'s predecessor in this traversal (lines 8 and 22).
\end{itemize}
	
	In case a block \textit{b} is not yet visited, it is marked both visited and active (line 10) and the algorithm is called recursively for each successor (lines 12 and 13).
	The back-propagated loops are added to \textit{b}'s loops (line 13) and vice versa (lines 14 and 15).
	After the recursive traversal of all successors is finished, \textit{b} is set to inactive (line 17) and depending on its loop header flag the set of loops which contain \textit{b}'s predecessor is returned (line 20 or 22).
	Eventually, the first block of the function is processed and at that point the set of loops has to be empty to ensure that the context of all loops has been closed correctly. 
	If this set is not empty, there is a path into some loop which does not pass the initially identified header yielding multiple loop entries and thus, irreducibility. \par
	We believe that irreducible control flow in unstructured languages is rare; in fact, we did not encounter it in the benchmarks used in the evaluation (see \Cref{par:benchmarks}).
	Thus, we omit the handling of irreducible control flow and bail out (i.e., we execute the 
	function without trying to reconstruct its loops) for functions that contain it.\par

\begin{figure*}[ht]
		\includegraphics[width = 0.9\textwidth]{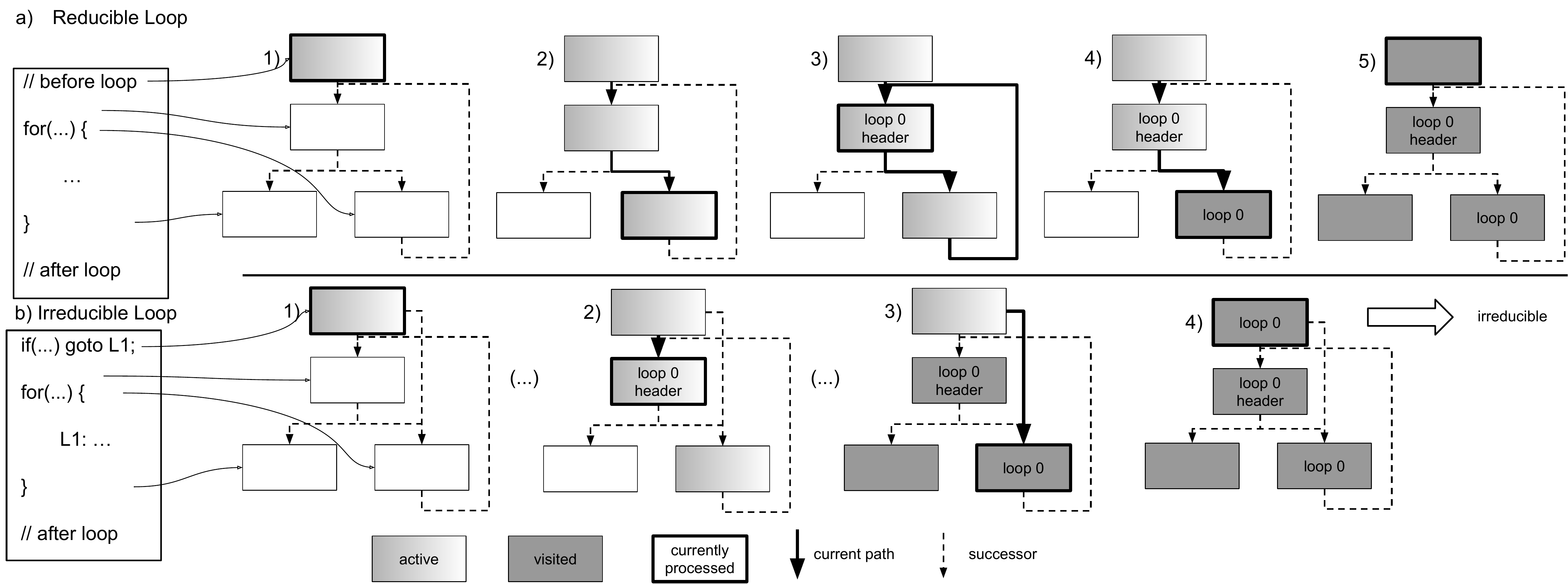}
		\caption{Visualization of two Loop Detection applications. Subfigure a) shows the detection of a reducible loop while subfigure b) shows an irreducible loop being detected as such, denoted by a loop ID set for the first block}
		\label{fig:loopDetection}
	\end{figure*}

	\Cref{fig:loopDetection} shows two examples of loop detection applications, one with a simple, reducible loop and the other with a slightly changed, yet irreducible control flow (having two loop entry points).
	In the reducible example, in Step 3 the backedge points to an active block which is marked as loop header and triggers back-propagation 	of the newly created loop ID.
	After Step 4 the left most block is processed and then in Step 5 backtracking arrives at the start block.
	Note, that some intermediate steps in the irreducible example are omitted as up to Step 3 everything is identical to the example with reducible control flow.
	In Step 3, the second entry to the loop (bypassing the header) is processed and the loop ID is thus propagated upwards to the start block.
	The irreducible loop is detected, because the function start block, which can never be part of a loop in LLVM IR, got a loop ID assigned. 
	
\subsection{Identifying Loop Relations} \label{ssec:loopRelations}
    We want to support OSR not only for the outermost loops, but also for all inner ones. 
    To achieve this, we need to model a loop nesting based on which we build a loop hierarchy in our node structure later on.
    Since the loop detection is unaware of the nesting level, we have to manually determine a contains-relationship between the loops. This is done by recursively checking if a loop contains the header of another loop, which is then added to its set of inner loops. 
    We use the depth-first approach to additionally have the loops sorted, that is, an outer loop is processed after its contained inner loops. 
    Hence, when creating \texttt{LoopNodes} we can use the established order to have all inner loops resolved by the time a dependent outer loop is created. \par
	Currently, we do not identify loop relations for irreducible loops.
	While they could be partially supported by adapting our approach (see \Cref{ssec:discussion}), it is still necessary to determine which loop is an inner loop and which is an outer loop.
	If this distinction is impossible, we are not able to establish the node hierarchy because inner loops have to be processed before the outer loop is created.
	Thus, we check for a bi-directional contains-relationships and bail out for such constructs. \par 

\subsection{Node Creation and Integration} \label{ssec:nodeCreation}
	To use Truffle's OSR mechanism, we create nodes for identified loops and embed their execution within the dispatch loops of basic blocks. 
	This includes implementing the Truffle \texttt{LoopNode} interface for modeling loops, adopting the block dispatch to also support high-level loop structures and providing a way for executing the loop. \par
\begin{figure}[ht]
		\includegraphics[width = 0.45\textwidth]{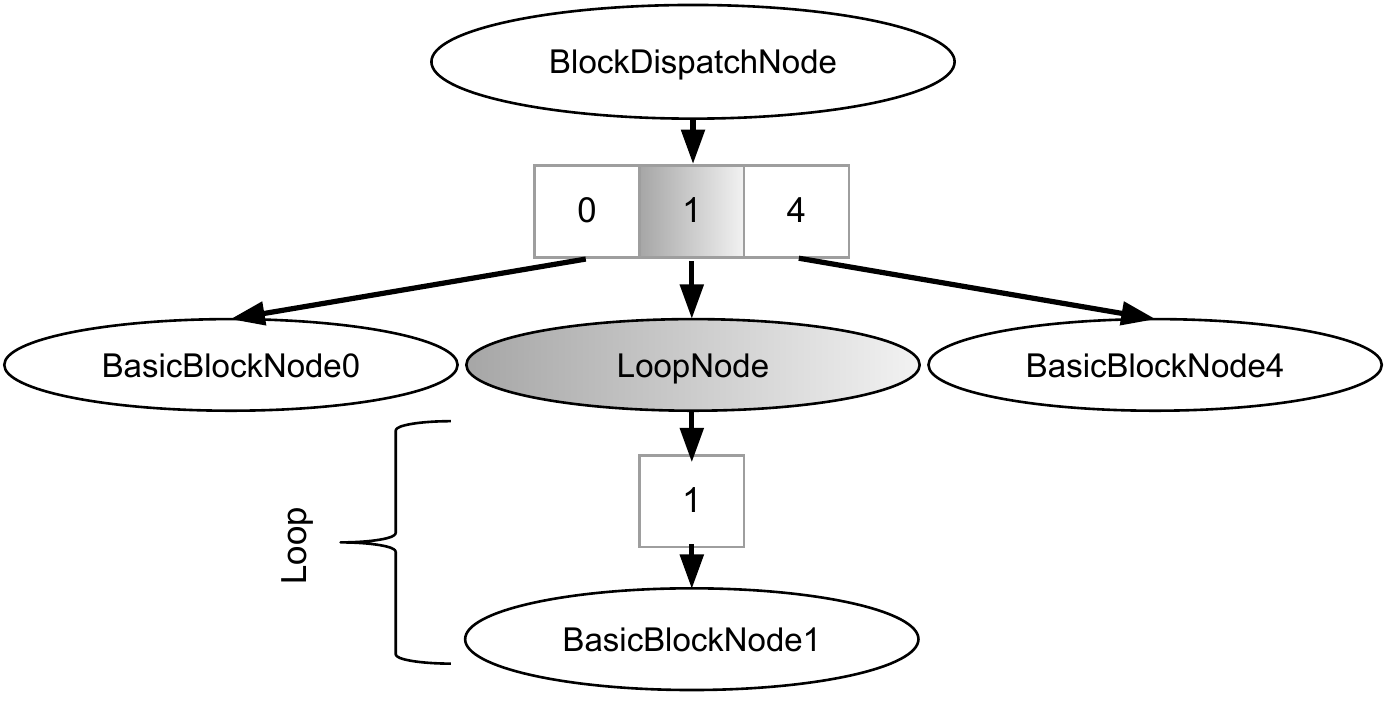}
		\caption{Node hierarchy; compacted for brevity (no \texttt{RepeatingNode} and \texttt{LoopDispatchNode} modeled)}
		\label{fig:nodeHierarchy}
	\end{figure}

	The function parser creates \texttt{LoopNodes} together with all other Truffle nodes when parsing a method. 
	After the control flow analysis, it wraps the \texttt{BasicBlockNodes} forming a loop into a \texttt{LoopDispatchNode}, which models the loop body and is responsible for executing the loop. 
	The \texttt{LoopDispatchNode} is then used to create \texttt{RepeatingNodes} and \texttt{LoopNodes} conforming to the Truffle interface. 
	Each \texttt{LoopNode} replaces the \texttt{BasicBlockNode} which denotes the header of the very same loop in the array of \texttt{BasicBlockNodes}.
	Thus, already created \texttt{LoopNodes} can be found in the body of outer loops. 
	To establish this node nesting, the function parser has to process loops iteratively from inside out, so that outer loops are created after their inner loops.
	However, the parser automatically ensures this by processing the loops according to the established order as described in \Cref{ssec:loopRelations}. 
	Finally, the function node is created by using the array of \texttt{BasicBlocks} and \texttt{LoopNodes} which are not already nested in another loop.
	This leads to a semi-hierarchical node structure, in comparison to the initially flat \texttt{BasicBlockNode} array. 
	\Cref{fig:nodeHierarchy} depicts this internal structure based on the function in \Cref{lst:problemSubFigA}. 

\subsection{Loop Execution}
	There are two ways to execute implicit unstructured LLVM IR loops as explicit loops in Sulong.
	Either by a full reconstruction the loop structure (body and condition nodes are not distinguished during loop detection), or by wrapping a block dispatch logic for the loop into the \texttt{LoopNode}.
	The latter approach can be justified by taking a look at it from the compiler's perspective. 
	In that sense, an OSR loop is a special call target which is optimized after several calls (i.e., iterations). 
	Therefore, the new \texttt{LoopDispatchNode} is closely related to a \texttt{BlockDispatchNode} for functions, yet with some differences.
	First, less terminating instructions have to be handled, because a \texttt{return} block, for example, can never be part of a loop as there is no path back to the loop header.
	Second, no function arguments have to be handled.
	Those two points make a \texttt{LoopDispatchNode} simpler than its counterpart for functions.
	However, a \texttt{LoopDispatchNode} has to return after each loop iteration or if the loop is exited.
	This check for ending an iteration is more costly than for ending a function. 
	Rather than dispatching to the next block until no successor is available (successor = -1; end of function), the loop dispatch has to both check if the successor is the loop header which indicates that the iteration has ended or if the loop is exited by dispatching to one of its successors. 
	However, Graal optimizes this additional checks away.\par
	While a simple loop condition is mapped to only one LLVM block and can thus be used directly as a condition for executing one loop iteration, more complex boolean conditions also consist of multiple blocks. 
	Therefore, a wrapper node would be needed, including another dispatch solely for loop conditions. 
	In order to maintain simplicity and due to the fact that it should be transparent whether a loop is exited because the condition failed or because another exit path was taken (e.g. \texttt{return}, \texttt{break}, \texttt{goto}, ...), the condition is pulled into the loop body. 
	In addition to having solved the problem of handling multi-condition loops, the distinction between condition and body nodes is made superfluous. \par
	
\subsection{Successor Determination}
	The final task is to continue function execution at the correct successor block after a loop is executed. 
	However, in contrast to structured control flow, our reconstructed loops can have arbitrary successors reached by jumps from within the loop. 
	Therefore, when entering a loop, it is not known in advance where to continue the execution after the \texttt{LoopNode} is evaluated. 
	While the \texttt{LoopDispatchNode} could return the successor at run time, the enclosing \texttt{RepeatingNode} has its predefined interface (see \Cref{lst:executeRepeating}) hindering passing the successor to the caller as return value. 
	However, we can use the run time stack for storing the successor value in a predefined, constant frame slot \texttt{LoopDispatchNode}. 
	While this process might seem costly, performance of tight loops is only affected in interpreted mode, due to frame virtualization in compiled programs.\par
	
\begin{figure}[ht]
		\includegraphics[width = 0.45\textwidth]{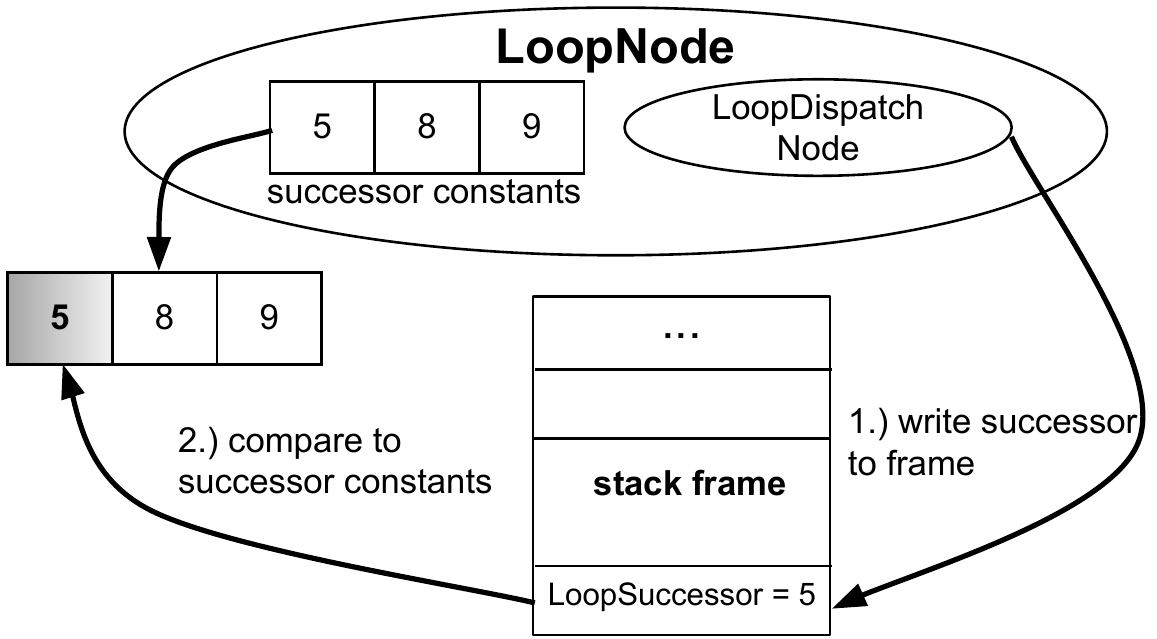}
		\caption{Successor Determination. The successor discovered at run time is written to a stack frame and used as a lookup into the set of constant successors.}
		\label{fig:successorDetermination}
	\end{figure}

	In order to generate code that can be easily optimized by Graal, the block dispatch loop (see \Cref{lst:interpreterloop}) and all block successors have to be constant (see \Cref{sec:PE}).
	Thus, when reading the loop successor from a run-time-written frame slot, the interpreter loop cannot be unrolled. 
	We solve this issue by storing the set of constant successors in each \texttt{LoopNode} and use a lookup loop for finding the constant successor with the same value as in the frame slot.
	Then, the constant value is used to determine the successor block.
	As the compiler can determine the set of possible successors for each loop, it can unroll the dispatch loop accordingly.
	This approach is visualized in \Cref{fig:successorDetermination}, where the value in the frame slot is shown to be used as a lookup into the set of constant successors.

\subsection{Irreducible Control Flow} \label{ssec:discussion}
	Graal can not handle irreducible control flow (ICF), but rather uses a dispatch loop for such patterns.
	A similar idea is used for our approach implemented in Truffle based languages, which is a way to have irreducible control flow reduced to one AST node.
	However, support for ICF could be added as outlined below.
    Irreducible loops are characterized by having multiple entries. 
    Thus, a loop cannot be collapsed to one black box block, as the entry point might differ from execution to execution. 
    There are approaches for transforming irreducible into reducible control flow, which mostly center around code duplication~\cite{Erosa1994}. \par
	
	We initially experimented with the Tarjan algorithm~\cite{Tarjan1971} for detecting strongly connected components in order to find such irreducible loops. 
	However, due to additional complexity for resolving nested loops, code growth for duplicated nodes and the fact that ICF was never encountered in our benchmarks this approach was discarded in favor of bailing out in case of ICF.
\section{Evaluation}
	We evaluated our approach which we prototyped in Sulong with a set of C benchmarks to 
	investigate improvements in warm-up and impacts on peak performance.
	
	\begin{figure*}[ht]
		\includegraphics[width = 0.8\textwidth]{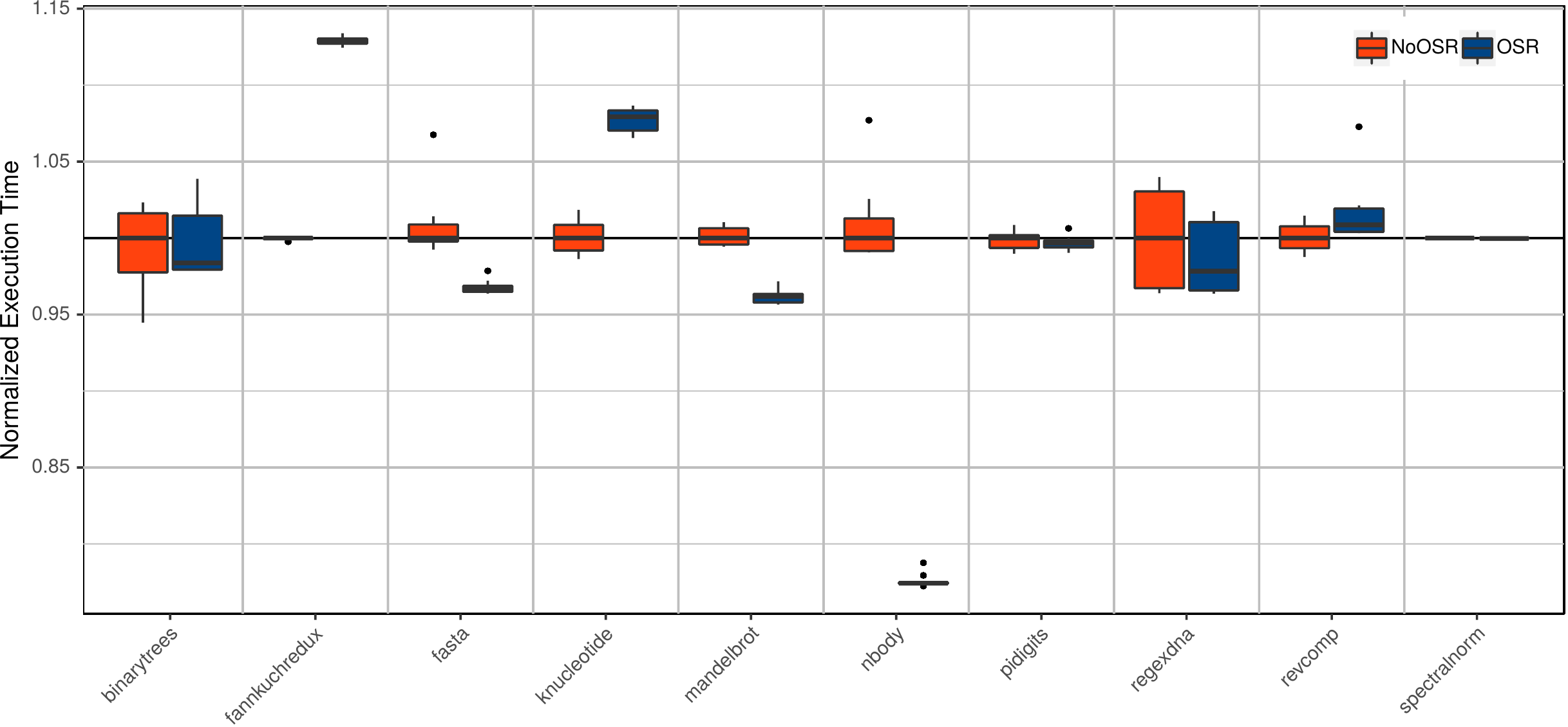}
		\caption{Normalized execution times after reaching peak performance of Sulong with and without OSR; lower is better.}
		\label{fig:peak_clbg}
	\end{figure*}
	
	\begin{figure*}[ht]
		\includegraphics[width = 0.8\textwidth]{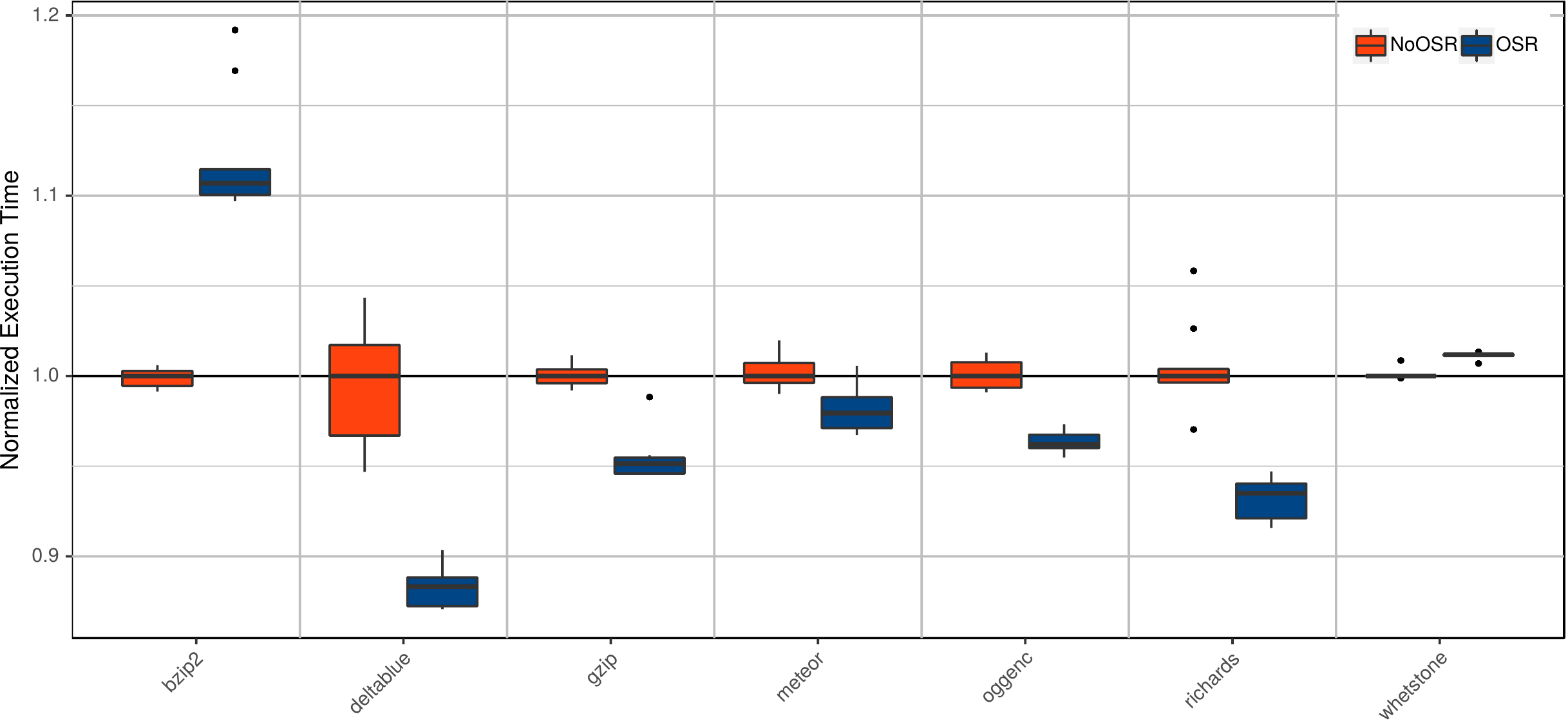}
		\caption{Normalized execution times after reaching peak performance of Sulong with and without OSR; lower is better.}
		\label{fig:peak_others}
	\end{figure*}
	
	\begin{figure*}[ht]
		\includegraphics[width = 0.8\textwidth]{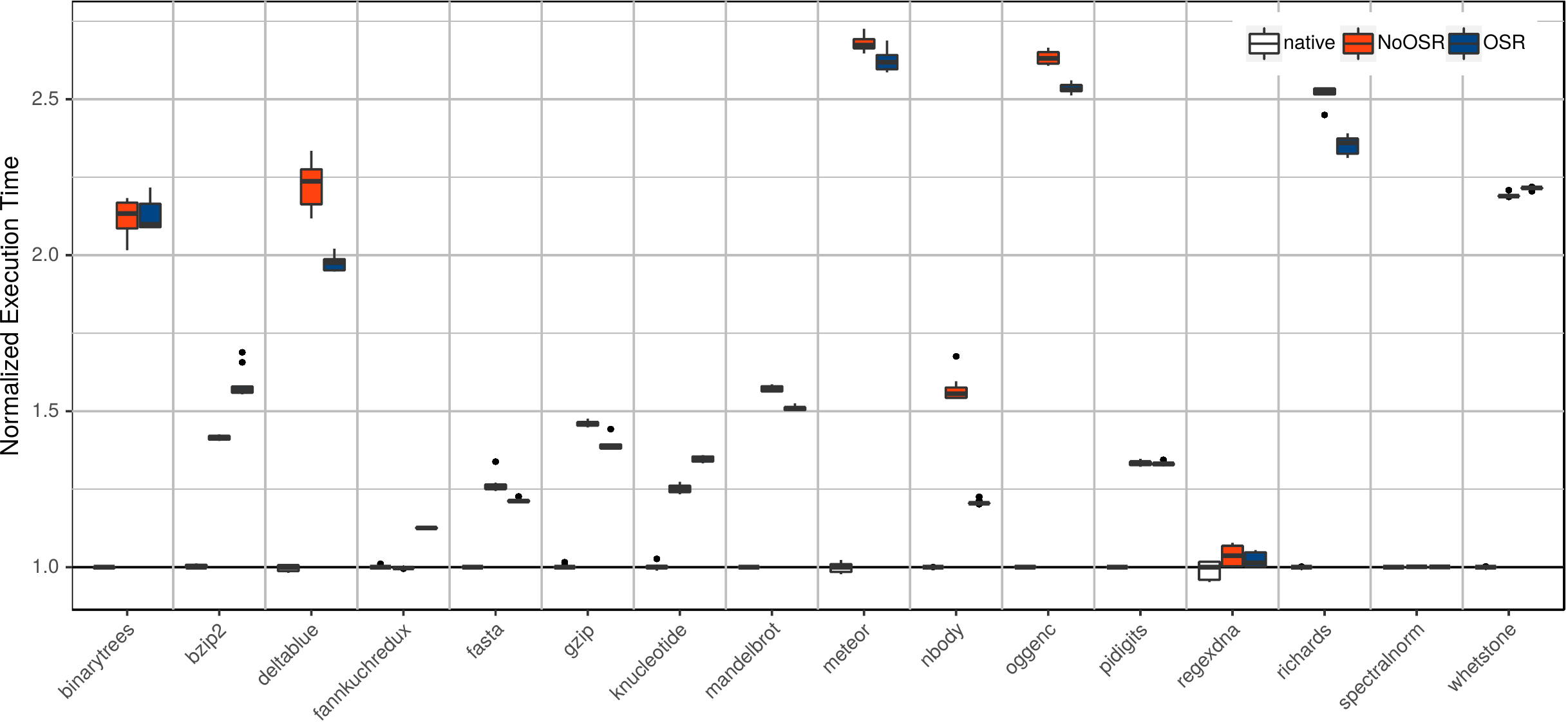}
		\caption{Normalized execution times after reaching peak performance of Sulong without OSR-related changes compared to native benchmark execution. Lower is better.}
		\label{fig:peak_native}
	\end{figure*}

\paragraph{Hypothesis}
	We hypothesize that enabling OSR reduces the program warm-up significantly while not influencing the peak performance. 

\subsection{Setup}

\paragraph{Benchmarks} \label{par:benchmarks}
	We primarily selected benchmarks from the \emph{Computer Language Benchmarks Game (CLBG)}\footnote{\url{https://benchmarksgame-team.pages.debian.net/benchmarksgame/}}. 
	They are micro-benchmarks and do not reflect the behavior of real-world applications. 
	However, their small size and their structure---benchmarks like fannkuch have long-running and computation-intensive loops---makes them suitable to study the benefits of OSR. 
	The problem sizes were chosen to allow for multiple benchmark runs. 
	In addition, we used other popular benchmarks like \emph{whetstone}\footnote{\url{www.netlib.org/benchmark/whetstone.c}}, \emph{deltablue}\footnote{\url{https://constraints.cs.washington.edu/deltablue/}} and \emph{richards}\footnote{\url{https://www.cl.cam.ac.uk/~mr10/Bench.html}}. 
	Furthermore, we evaluated our approach with three larger applications, \emph{bzip2}, 
	\emph{gzip} and \emph{oggenc}, which are part of the \emph{Large single compilation-unit C 
	programs}\footnote{\url{http://people.csail.mit.edu/smcc/projects/single-file-programs/}}, 
	to better investigate the impact on real world programs. 

\paragraph{Harness}
	We used a configurable harness for evaluating the warm-up behavior and peak performance of the benchmarks and conducted ten out-of-process benchmark executions with at least 50 in-process iterations.
	We analyzed warm-up behavior by analyzing the first ten in-process iterations of the respective benchmark.
    We verified that within these ten iterations, execution of the benchmarks reached peak performance.
	For peak performance, we evaluated only the last ten in-process iterations of each benchmark.

\paragraph{Environment}
	Our benchmarking machines are equipped with Intel Xeon CPU E5-2699 processors with 72 cores 
	at 2.30GHz each along with 256GB of RAM.
	For compiling the benchmarks to LLVM IR we used the LLVM front end Clang 3.8 at optimization level 03. 
	Sulong is executed on top of GraalVM, including the default Truffle OSR threshold of 100 000 iterations. 
	We compare Sulong with the implemented OSR approach against its version without any OSR related changes. 
	
\subsection{Results} \label{sec:results}
	
	\begin{figure*}[ht]
		\includegraphics[width = 0.83\textwidth]{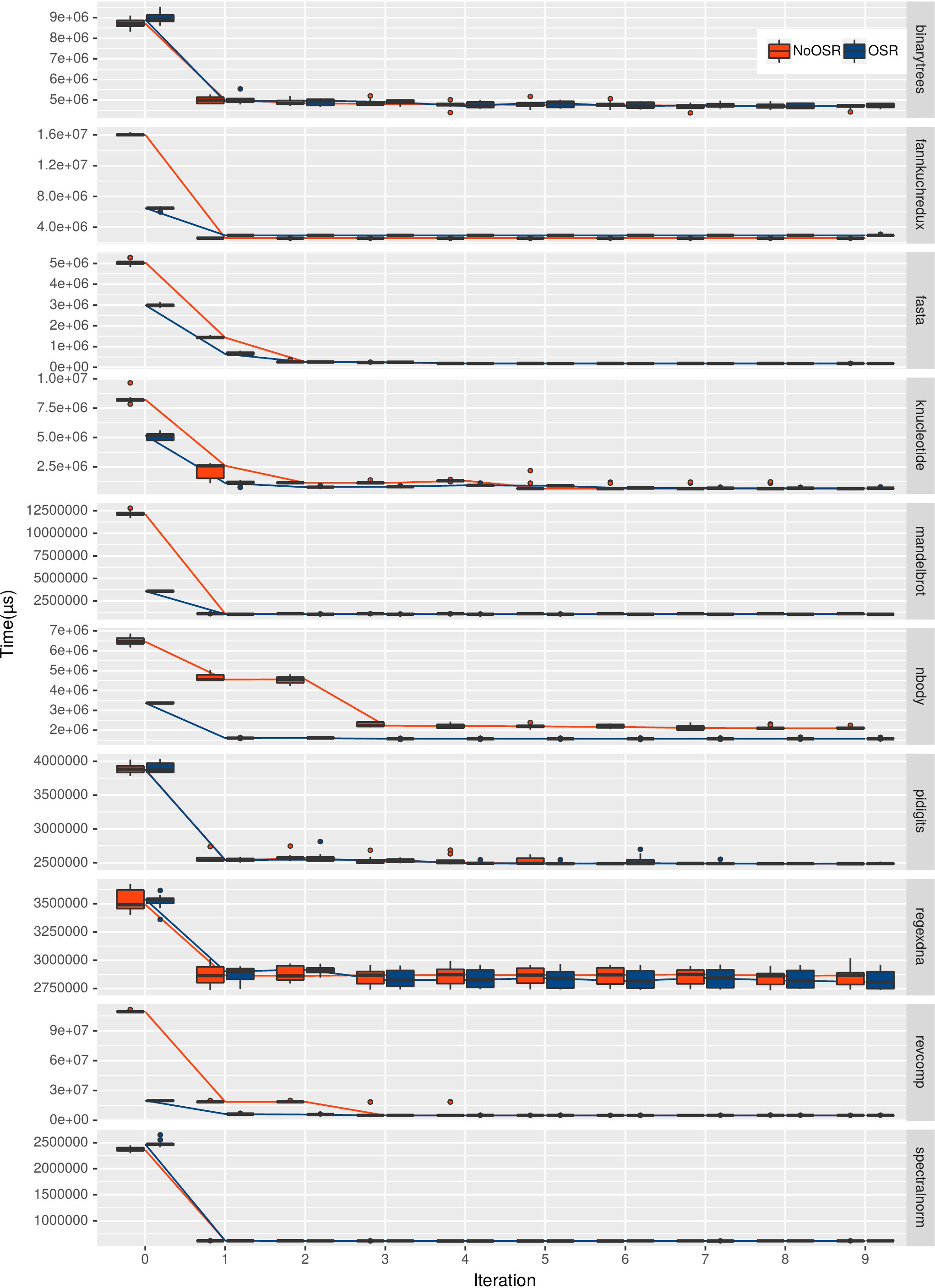}
		\caption{Warm-up of Sulong with and without OSR; the x-axis shows the sequence number in the series of consecutive in-process benchmark executions, while the y-axis shows the execution time of a benchmark execution. Lower is better.}
		\label{fig:warmup_clbg}
	\end{figure*}
	
	\begin{figure*}[ht]
		\includegraphics[width = 0.83\textwidth]{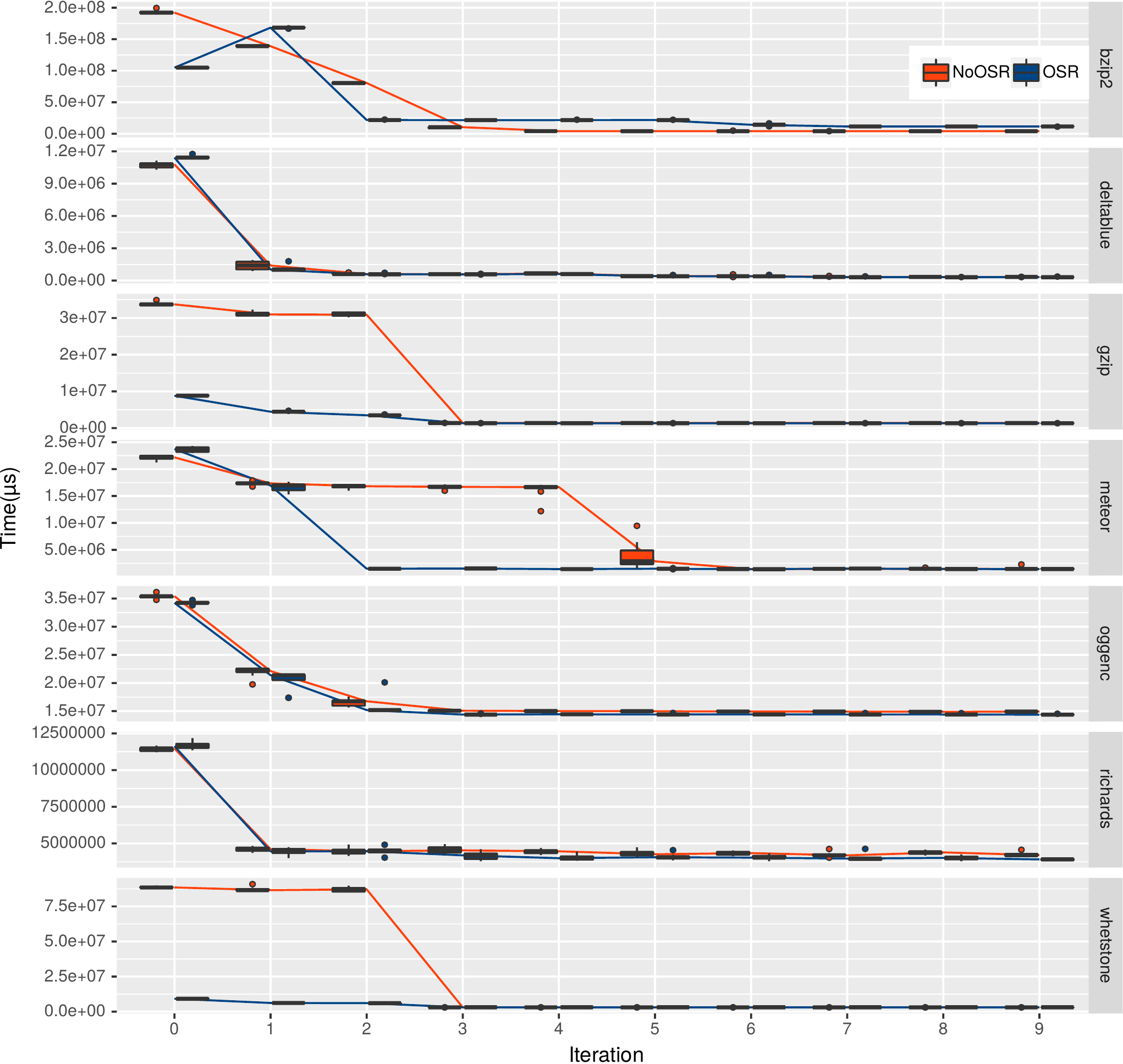}
		\caption{Warm-up of Sulong with and without OSR; the x-axis shows the sequence number in the series of consecutive in-process benchmark executions, while the y-axis shows the execution time of a benchmark execution. Lower is better.}
		\label{fig:warmup_others}
		\vspace{-8pt}
	\end{figure*}
	
	\paragraph{Native Execution}In order to show that Sulong is a mature execution system, we compared its performance against state of the art compiled native executables.
	\Cref{fig:peak_native} gives a peak performance comparison of the benchmarks executed natively and using Sulong with and without OSR related changes.
	Warm-up is of course not present in the native executions and thus omitted to be shown.
	Although no contribution proposed in this paper is visible in these benchmark runs, we want to provide a context for the peak performance deviations imposed by our approach.
	Most benchmarks are, when executed on Sulong, slower by factors up to 2.8x with 
	\emph{revcomp} being about 6x slower.
	Thus, \emph{revcomp} is not visualized to keep the plot readable.
	\Cref{tab:peakPerformance} contains a condensed depiction of the performance data including the median and the standard error for each benchmark.	
	
	\begin{table}[h!]
		\caption{Overview of peak performance for benchmarks normalized to their native execution and the standard error (SE). Lower is better.}
		\resizebox{0.48\textwidth}{!}{
			\begin{tabular}{|l|ll|ll|ll|}
				\hline
				\multirow{2}{*}{} & \multicolumn{2}{c|}{native} & \multicolumn{2}{c|}{no OSR} & \multicolumn{2}{c|}{OSR}   \\
				& median     & SE & median     & SE & median    & SE \\ \hline
				binarytrees       & 1.0  & 2.449e-04   & 2.134  & 4.914e-03   & 2.099 & 4.832e-03   \\ \hline
				bzip2             & 1.0  & 4.645e-04   & 1.417  & 6.567e-04   & 1.568 & 4.435e-03   \\ \hline
				deltablue         & 1.0  & 9.799e-04   & 2.237  & 6.649e-03   & 1.976 & 2.379e-03   \\ \hline
				fannkuchredux     & 1.0  & 3.272e-04   & 0.997  & 8.543e-05   & 1.126 & 2.459e-04   \\ \hline
				fasta             & 1.0  & 1.354e-04   & 1.254  & 2.618e-03   & 1.212 & 5.316e-04   \\ \hline
				gzip              & 1.0  & 5.093e-04   & 1.460  & 9.248e-04   & 1.389 & 1.745e-03   \\ \hline
				knucleotide       & 1.0  & 9.804e-04   & 1.251  & 1.340e-03   & 1.350 & 9.148e-04   \\ \hline
				mandelbrot        & 1.0  & 3.973e-05   & 1.570  & 8.501e-04   & 1.510 & 6.628e-04   \\ \hline
				meteor            & 1.0  & 1.512e-03   & 2.673  & 2.196e-03   & 2.618 & 3.150e-03   \\ \hline
				nbody             & 1.0  & 1.187e-05   & 1.556  & 3.897e-03   & 1.205 & 6.787e-04   \\ \hline
				oggenc            & 1.0  & 2.314e-04   & 2.631  & 1.943e-03   & 2.531 & 1.446e-03   \\ \hline
				pidigits          & 1.0  & 8.932e-05   & 1.336  & 7.600e-04   & 1.331 & 5.900e-04   \\ \hline
				regexdna          & 1.0  & 2.610e-03   & 1.037  & 3.110e-03   & 1.014 & 2.335e-03   \\ \hline
				revcomp           & 1.0  & 6.581e-04   & 6.090  & 4.951e-03   & 6.143 & 1.219e-02   \\ \hline
				richards          & 1.0  & 7.472e-05   & 2.525  & 5.535e-03   & 2.360 & 2.816e-03   \\ \hline
				spectralnorm      & 1.0  & 1.559e-05   & 1.001  & 6.512e-06   & 1.001 & 9.955e-06   \\ \hline
				whetstone         & 1.0  & 8.373e-05   & 2.190  & 5.841e-04   & 2.216 & 3.595e-04   \\ \hline
		\end{tabular}}
		\label{tab:peakPerformance}
	\end{table}
	
	\paragraph{Peak-performance Impact}While OSR should conceptually not impact a program's peak performance, our experiments disproved this initial hypothesis.
	\Cref{fig:peak_clbg} and \Cref{fig:peak_others} show deviations with up to 13\% overhead (median) for \emph{fannkuch} down to 23\% speed-up (median) for \emph{nbody}.
	Only for three benchmarks (\emph{fannkuch}, \emph{knucleotide} and \emph{bzip2}) the peak performance is significantly worse (median at least 5\% worse).
	Three benchmarks (\emph{deltablue}, \emph{nbody} and \emph{richards}) show significantly higher peak performance (median at least 5\% better).
	The majority of our benchmarks, that are 11 out of 17, differ only slightly from the non-OSR version of Sulong, however with a tendency to increased peak performance.
	We found that due to the changes in the Truffle node structure, optimizations are applied differently by the Graal compiler, resulting in the observed deviations.
	It is difficult to attribute the peak performance changes to individual optimizations, as multiple overlapping compilation paradigms produce the measured results.
	
	\paragraph{Program Warm-up}We identified two categories of behaviors in the investigated benchmarks regarding program warm-up, which can be seen in \Cref{fig:warmup_clbg} and \Cref{fig:warmup_others}.
	Firstly, for applications with long-running loops, which trigger OSR, speed-ups by factors of 1.5x (\emph{knucleotide}) up to 9x (\emph{whetstone}) are encountered for the first one to four iterations.
	This can be seen for benchmarks \emph{fannkuch}, \emph{fasta}, \emph{knucleotide}, \emph{mandelbrot}, \emph{nbody} and \emph{revcomp (reverse-complement)} in the CLBG suite in \Cref{fig:warmup_clbg} and also for \emph{gzip}, \emph{meteor} and \emph{whetstone} in \Cref{fig:warmup_others}. 
	The warm-up of \emph{bzip2} shown in \Cref{fig:warmup_others} behaves differently as the second iteration shows a peak in execution time, which is linked to a de-optimization issue tied to this benchmark.
	Secondly, there are benchmarks where the program warm-up is not affected by OSR at all.
	This is either the case in the absence of loops or if loops do not reach the iteration threshold for OSR compilation.
	Minor deviations in warm-up result from the changes in the node structure.
	
\subsection{Discussion}
The results suggest that our OSR approach can significantly reduce warm-up time.
The actual improvement is highly dependent on the problem size of the benchmark, because for larger problem sizes, loops are often running longer in interpreted mode.
For example, in \emph{fannkuch}, the problem size is the length of a permutation array, which determines the number of loop iterations needed to cover all possible permutations.
With a problem size of 11, \emph{fannkuch} did not finish within a day without OSR, but in less than 30 minutes with our approach enabled.
For evaluation feasibility \emph{fannkuch's} problem size was reduced to 9.
The evaluation also demonstrated that peak performance is affected by our approach.
Due to changes in the Truffle node structure, a different program is produced which explains the deviations.
Some benchmarks show peak performance regressions, others speed up, however with a tendency to the latter.
\section{Related Work}

To the best of our knowledge, we have presented the first approach for providing OSR in a Truffle implementation for an unstructured language.
Below, we consider the wider context of related work.

\paragraph{OSR}
On-stack replacement was first researched by ~\citet{Holzle94} as a strategy to switch between different versions of an executed method in the context of (re-)compilation and dynamic de-optimization~\cite{Holzle92} to support debugging of optimized code.
OSR is supported by most of the popular method-based dynamic compilers.
For example, the HotSpotVM~\cite{HotSpot}, which is one of the most popular VMs for Java bytecode, supports OSR by incrementing counters at loop backedges that trigger compilation when a counter overflows.
When compilation finishes, the next executed backedge will transition to the compiled version of the loop.
Values in the interpreter frame are mapped to an OSR buffer, from which the compiled code extracts the values.
To support compilers in applying OSR for unstructured languages we proposed reconstructing and extracting loops.
\Citet{DElia2016} conducted a case study on how to implement function-level on-stack replacement for LLVM at arbitrary points, by using glue code to facilitate a smooth transition between the two versions of a function. 
In their recent work they focused on a more abstract view on OSR in the context of code transformation~\cite{DElia2018}.
In~\cite{DElia2018} a more general way to apply OSR is proposed by enabling transitions at any code location.

\paragraph{Truffle Bytecode vs. AST Interpreters}
\citet{Niephaus18} compared an AST-based with a bytecode-interpreter-based Truffle implementation for Squeak/Smalltalk. 
For the AST-based approach, they had to decompile the generated bytecode to construct high-level Truffle AST nodes.
They also report significant performance gains when using Truffle's \texttt{LoopNode} interface, which required additions in the decompilation process, to successfully reconstruct loop condition and body.
They observed a warm-up period for the AST-based interpreter, but not for the bytecode interpreter approach.
However, they remarked that the results might not be generalizable, since they evaluated their implementation with only two micro-benchmarks.
We speculate that the bytecode-based version of Squeak/Smalltalk, would also benefit from enabling OSR like in Sulong.

\paragraph{Reconstruction of loops}
Loop reconstruction is a well researched topic, which dates back into the 1970s with Tarjan~\cite{Tarjan1973, Tarjan1971} formulating his interval analysis algorithm capable of identifying loops in reducible control flow graphs.
The algorithm creates a depth-first tree of the CFG and identifies loops in a bottom up traversal from inside out, by collapsing inner loops into single vertices~\cite{Ramalingam1999}.
This depth-first nature is found in many loop reconstruction algorithms being developed over the years---including the one used in this paper.
Havlak refined Tarjan's algorithm to work with irreducible control flow too~\cite{Havlak1997}, however, in quadratic time~\cite{Ramalingam1999}.
But Havlak's algorithm is still used in more recent work~\cite{Sato2011}.
Ramalingam then extended Havlak's algorithm to identify both reducible and irreducible loops in almost linear time~\cite{Ramalingam1999}.
Other algorithms build on the work of Tarjan and add support for irreducible control flow~\cite{Steensgaard1993, Sreedhar1996}, which slightly differ in the returned irreducible loops~\cite{Ramalingam1999}.
For our work, as irreducible control flow is not supported, the simple depth-first reconstruction of reducible loops in combination with the detection of irreducible loops is sufficient.

\paragraph{Reconstruction of high-level control flow}
\citet{Leopoldseder2015} described an approach to reconstruct high-level language constructs from compiler IR.
They compiled Java bytecode to JavaScript by first translating it to Graal IR~\cite{Duboscq13a} and then reconstructing high-level constructs.
They faced similar problems like we did, for example, the problem of handling different successor blocks of loop exits.
However, their approach merged loop exits whereas we use the dynamically identified loop successor as lookup for the constant successors, known at compile time. \par
Zakai~\cite{Zakai2011} introduced the \emph{Emscripten} compiler, which translates LLVM IR to JavaScript.
In its architecture, a module for reconstructing high-level JavaScript loops from LLVM IR is presented, called \emph{The Relooper}.
They also point out that due to extensive use of \texttt{goto} statements no meaningful high-level structure might be re-established.

\section{Conclusion and Future Work}
OSR for loops in unstructured languages is problematic due to the lack of high-level representations on which optimizations can be performed.
In this paper, we have presented an approach for enabling OSR in Truffle-based interpreters for unstructured languages by partially reconstructing high-level loops from basic blocks.
Unlike traditional OSR, we wrap loops into AST nodes to enable Truffle to extract the \textit{LoopNodes} into separate \textit{CallTargets}, which are function equivalents and can be OSR-compiled after each loop iteration.
We implemented this approach in Sulong, and demonstrated that it can significantly reduce warm-up time.
Our approach is applicable to any other Truffle based bytecode interpreter, but also other languages can implement similar approaches building on our multi-tier system.

As part of future work, instead of bailing out on whole functions when irreducible control flow is detected, reducible loops could be handled correctly while irreducible loops are not detected as loops at all. 
This would be useful for functions with many loops where few isolated irreducible loops would then not prevent OSR compilation for the others as well.
Alternatively, support for irreducible loops could be implemented for Sulong as suggested in \Cref{ssec:discussion}.
For a more detailed evaluation, the used benchmarks could be analyzed and compared in terms of their code structure (number of loops, call sites, etc.).

\bibliography{biblio}

\end{document}